\documentstyle[a4wide,12pt]{article}
\title{Reply to ``The quantum algorithm of Kieu does not 
solve the Hilbert's tenth problem"}
\author{Tien D Kieu~\footnote{Centre for Atom Optics and Ultrafast Spectroscopy, Swinburne
University of Technology, Hawthorn 3122, Australia}~\footnote{email: kieu@swin.edu.au}}
\begin{document}
\maketitle
\begin{abstract}
The arguments employed in {\tt quant-ph/0111009}, to claim that the quantum
algorithm in {\tt quant-ph/0110136} does not work, are so general that were they true
then the adiabatic theorem itself would have been wrong.  As a matter of
fact, those arguments are only valid for the sudden approximation, not
the adiabatic process.
\end{abstract}
The author of~\cite{tsirelson} carefully distinguishes between the general 
ground-state oracle from the algorithm which  explicitly employs the adiabatic 
evolution, both proposed for the Hilbert's tenth problem in~\cite{kieu}.  
Then it is concluded that this latter quantum algorithm is untenable.  
However, the
arguments employed to reach this conclusion is so general.  They are
apparently applicable not only to the quantum algorithm but also to any
adiabatic process.  Were they true then the adiabatic theorem would have
been wrong.  In the below we examine the crucial steps in the arguments
and point out their shortcoming.

We follow the notations of~\cite{tsirelson} and just pick up at the
crucial inequality (the un-numbered, last inequality of the paper)
\begin{eqnarray}
\||g(T)\rangle - |g_0(T)\rangle\| \le T\|
H_P|g_I\rangle\|.
\label{3}
\end{eqnarray}
where $|g(T)\rangle$ is the end state arrived at {\em some} time $T$ in a supposedly adiabatic
process which starts with the initial state $|g_I\rangle$ and ends with the hamiltonian $H_P$.  
The state $|g_0(T)\rangle$ is constructed so that it only differs from the
initial state $|g_I\rangle$ by a $T$-dependent phase factor.
Then from the fact that
\begin{eqnarray}
\lim_{x_{\rm min}\to \infty} \|
H_P|g_I\rangle\|
= \lim_{x_{\rm min}\to \infty} |\langle x_{\rm min} | g_I \rangle|  = 0
\label{4}
\end{eqnarray}
(where $|x_{\rm min} \rangle$ is the sought-after state, contained in
$H_P$), it was concluded that the left hand side of~(\ref{3}) can be
vanishingly small and thus that $|g(T)\rangle$ can never be closed to
$|x_{\rm min} \rangle$ for large $x_{\rm min}$.  Hence ``the adiabatic 
evolution fails," not just only the
proposed quantum algorithm for the Hilbert's tenth problem.

But this is not the case.

All that~(\ref{3}) could say is if $T$ is not
long enough then the adiabatic approximation cannot be applied.  As a matter 
of fact, for sufficiently short $T$ (as compared to the inverse of the
convergence in~(\ref{4})) the inequality~(\ref{3}) is just a
statement of the sudden approximation, where the state does not change
appreciately over that period $T$.

The point that is missed by the author of~\cite{tsirelson} is that $T$
cannot be
{\it fixed} for all processes, in contrast to his phrases ``... during a time $T$ 
that does not depend on $P$" and ``... $x_{\rm min}$, unknown to the user,
does not influence $T$, $H_I$ and $g_I$."  

Depending on the problem (i.e. the initial and final
hamiltonians involved) 
the adiabatic time $T$ has to be accordingly long.
A single $T$ cannot 
solve all the problems.  
This time is the counterpart of the computational complexity of an
algorithm whose number of steps varies according to the size of the input.
For each problem we have to (over)estimate the 
appropriate $T$ through an estimate of the energy gap between the
ground and first excited states.  The adiabatic theorem stipulates that
the time $T$ should be much larger than the inverse of this gap for a high 
probability in the instantaneous ground state.  
This dependence of $T$ is such that the product on the right hand 
side of the last inequality,
$ T\times|\langle x_{\rm min} | 
g_I \rangle |$, should not be vanishingly small.

In other words, the key point in applying the adiabatic theorem is to 
estimate the evolution time $T$ for the particular problem at hand and run the
algorithm over that period to get a reasonable success probability.
Without knowing exactly what $|x_{\rm min}\rangle$ is,
the adiabatic theorem still gives some estimate of
$T$ through the size of the gap, of which an estimation method is 
proposed in~\cite{kieu}.  Different Diophantine equations lead to
different gaps (which also depend on the initial hamiltonians $H_I$).  
Provided this gap can be arranged, through the available 
freedom mentioned in~\cite{kieu}, not to be identically zero at any time 
then the required $T$ will be {\it finite}.  Also contained in~\cite{kieu} 
are some discussions on the verification of the estimated $T$ and of
the results obtained at the end of the quantum computation.  Numerical 
study of some cases of interests will be presented elsewhere.

It may appear from~(\ref{3}) that a quantum machine can only explore a
finite domain in a finite time and is thus no better than a classical
machine in terms of computability.  But there is a crucial
difference.

In a classical search even if the global minimum is come across, it cannot 
generally be 
proved that it is the global minimum (unless it is a zero of the Diophantine 
equation).  Armed only with mathematical logic, we would still have to 
compare it with all other numbers from the 
infinite domain yet to come, but we obviously can never complete this 
comparison in finite time --thus, noncomputability.

In the quantum case, the global minimum is encoded in the ground
state.  Then, by energetic tagging, the global minimum can be found in finite 
time and confirmed, if it is the ground state that is obtained at the 
end of the computation.  And the ground state can be indentified and/or verified by 
physical principles.  These principles are over and above 
the mathematics which govern the logic of a classical machine and help
differentiating the quantum from the classical.  Quantum mechanics could 
``explore" an infinite domain, but only in the sense that it can select, 
among an infinite number of states, one single state (or a subspace in case 
of degeneracy) to be identified as the ground state of some given hamiltonian
(which is bounded from below).  
This ``sorting" can be done because of energetic reason, which is a physical 
principle and is not available to classical computability.  

Information {\it is} physical, after all.

\section*{Acknowledgement}
I wish to thank Falk Scharnberg for discussion and Boris Tsirelson for 
email correspondence.

\end{document}